\documentclass[11pt,prd,onecolumn,showpacs,amsmath,amssymb,aps,floats,floatfix]{article}
\usepackage[flushleft]{threeparttable}
\usepackage{authblk}
\usepackage[sort,numbers]{natbib}
\usepackage{hyperref}
\usepackage{comment}
\usepackage{dcolumn}
\usepackage{bm}
\usepackage{amsmath}
\usepackage{amsfonts}
\usepackage{amssymb}
\usepackage{color}
\usepackage{latexsym}
\usepackage{slashed} % slash mark
\usepackage{pstricks}
\usepackage{indentfirst}
\usepackage{mathrsfs}
\usepackage{multirow}
\usepackage{epsfig,psfrag}
\usepackage{subfigure}
\usepackage{setspace} % spacing

\usepackage{amsmath}
\graphicspath{{finalpic/}}  % figure path \right\}
\usepackage{float}
\usepackage{footnote}

\usepackage{indentfirst}

\usepackage{graphicx}
\usepackage{epsfig,subfigure,psfrag}
\usepackage{mathrsfs}

\usepackage{slashed} % slash mark

\usepackage{epstopdf}

\def \be {\begin{equation}}
\def \ee {\end{equation}}
\def \ba {\begin{eqnarray}}
\def \ea {\end{eqnarray}}

\def\TeV{\mathrm{TeV}}     % TeV
\def\GeV{\mathrm{GeV}}     % GeV
\def\MeV{\mathrm{MeV}}     % MeV
\def\kpc{\mathrm{kpc}}
\makeatletter
\newcommand*{\rom}[1]{\expandafter\@slowromancap\romannumeral #1@}
\makeatother

\newcommand {\dbar}{{d\kern-.22em\lower-.73ex\hbox{-}}}

\newcommand {\fslash}[1]{{#1\kern -0.6em / \kern 0.2em}}

\def\erg{\mathrm{erg}}
\def\kyr{\mathrm{kyr}}
\def\ev{\mathrm{eV}}
\def\log{\mathrm{log}}
\def\exp{\mathrm{exp}}
\def\sr{\mathrm{sr}}
\def\kpc{\mathrm{kpc}}
\def\GV{\mathrm{GV}}
\def\MV{\mathrm{MV}}
\def\cm{\mathrm{cm}} % cm
\def\km{\mathrm{km}} % km
\def\sec{\mathrm{s}} % s
\def\TeV{\mathrm{TeV}} % TeV
\def\GeV{\mathrm{GeV}} % GeV
\def\MeV{\mathrm{MeV}} % MeV
\makeindex

\begin{document}
\title{Explanations of the DAMPE high energy electron/positron spectrum in the dark matter annihilation and pulsar scenarios}
\author[1,2]{Bing-Bing Wang}
\author[1]{Xiao-Jun Bi}
\author[1]{Su-Jie Lin}
\author[1]{Peng-fei Yin}
\affil[1]{Key Laboratory of Particle Astrophysics, Institute of High Energy Physics, Chinese Academy of Sciences, Beijing 100049, China}
\affil[2]{University of Chinese Academy of Sciences, Beijing, 100049, China}

\maketitle

%\begin{landscape}
%\begin{minipage}{1\columnwidth}%
%\begin{minipage}{1\columnwith}%

\begin{abstract}
Many studies have shown that either the nearby astrophysical source or dark matter (DM) annihilation/decay is required to explain the origin of high energy cosmic ray (CR) $e^\pm$, which are measured by many experiments, such as PAMELA and AMS-02.
Recently, the Dark Matter Particle Explorer (DAMPE) collaboration has reported its first result of the total CR $e^\pm$ spectrum from $25 \,\GeV$ to $4.6 \,\TeV$ with high precision.
In this work, we study the DM annihilation and pulsar interpretations of the DAMPE high energy $e^\pm$ spectrum.
In the DM scenario, the leptonic annihilation channels to $\tau^+\tau^-$, $4\mu$, $4\tau$, and mixed charged lepton final states can well fit the DAMPE result, while the $\mu^+\mu^-$ channel has been excluded.
In addition, we find that the mixed charged leptons channel would lead to a sharp drop at $\sim$ TeV. However, these DM explanations are almost excluded by the observations of gamma-ray and CMB, unless some complicated DM models are introduced.
In the pulsar scenario, we analyze 21 nearby known pulsars and assume that one of them is the primary source of high energy CR $e^\pm$.
Considering the constraint from the Fermi-LAT observation of the $e^\pm$ anisotropy, we find that two pulsars are possible to explain the DAMPE data. Our results show that it is difficult to distinguish between the DM annihilation and single pulsar explanations of high energy $e^\pm$ with the current DAMPE result.
\end{abstract}

\section{Introduction} \label{section_introduction}

After the Cosmic Ray (CR) electron/positron excess above $10\GeV$ was confirmed by PAMELA and AMS-02 with high precision \cite{adriani2009anomalous,aguilar2014electron}, many studies on its origin have been proposed in the literature.
Two kinds of interpretations, including dark matter (DM) annihilation/decay in the Galactic halo \cite{bergstrom2008new,barger2009pamela,cirelli2009model,yin2009pamela,zhang2009discriminating,ArkaniHamed:2008qn,Pospelov:2008jd} and nearby astrophysical sources \cite{yuksel2009tev,hooper2009pulsars,profumo2012dissecting,malyshev2009pulsars,blasi2009origin,hu2009e+}, are widely studied.
Although the measurements of AMS-02 are unprecedentedly precise, those results are not yet sufficient to distinguish the two explanations~\cite{lin2015quantitative}.

The DArk Matter Particle Explorer (DAMPE) satellite launched on Dec.17, 2015 is a multipurpose detector, which consists of a Plastic Scintillator strip Detector (PSD), a Silicon-Tungsten tracker-converter (STK), a BGO imaging calorimeter, and a Neutron Detector (NUD).
Comparing to the AMS-02 experiment, DAMPE has a better energy resolution and could measure CR electrons and positrons at higher energies up to $10\,\TeV$.
Recently, the DAMPE collaboration reported its first result of the total CR $e^\pm$ spectrum in the energy range from 25~$\GeV$ to 4.6~$\TeV$~\cite{Ambrosi:2017wek}.
Many studies have been performed to explain the tentative features in this spectrum  ~\cite{Yuan:2017ysv,Fang:2017tvj,Fan:2017sor,Duan:2017pkq,Athron:2017drj,Cao:2017ydw,Liu:2017rgs,Huang:2017egk,Niu:2017hqe,Gao:2017pym,Yang:2017cjm,Cao:2017sju,Ghorbani:2017cey,Nomura:2017ohi,Gu:2017lir,Zhu:2017tvk,Li:2017tmd,Chen:2017tva,Jin:2017qcv,Duan:2017qwj,Zu:2017dzm,Ding:2017jdr,Gu:2017bdw,Chao:2017yjg,Tang:2017lfb,Gu:2017gle,Liu:2017obm,Ge:2017tkd}.

In this work, we explain the DAMPE $e^\pm$ result in the DM annihilation and pulsar scenarios, and perform a fit to the DAMPE $e^\pm$ spectrum and the AMS-02 positron fraction. The tentative feature at $\sim 1.4$ TeV in the DAMPE spectrum is not considered in this analysis.
Several leptonic DM annihilation channels to $\mu^+\mu^-$, $\tau^+\tau^-$, $4\mu$, $4\tau$, and mixed charged lepton final states $e^+e^-+\mu^+\mu^-+\tau^+\tau^-$ (denoted by $e\mu\tau$ for simplicity) are considered.
Although all these channels except for the $\mu^+\mu^-$ channel could provide a good fit to the DAMPE data, they would be excluded by the observations of gamma-ray and CMB in ordinary DM models.

For the pulsar scenario, we investigate 21 nearby pulsars in the ATNF catalog~\cite{manchester2005australia}\footnote{\url{http://www.atnf.csiro.au/people/pulsar/psrcat/}} and find out some candidates that can be the single source of the high energy $e^\pm$.
Furthermore, since the nearby pulsars may lead to significant anisotropy in the $e^\pm$ flux, we also adopt the anisotropy measurement of Fermi-LAT to explore the pulsar origin of high energy $e^\pm$, and find that some pulsars accounting for the DAMPE data could evade the current anisotropy constraint. Since the best-fit spectra resulted from some pulsars are very similar to those induced by DM annihilation, our results show that it is difficult to distinguish between these two explanations of high energy $e^\pm$ with the current DAMPE result.

This paper is organized as follows.
In Section \ref{section_electron_(positron)_propagation_in_galaxy}, we introduce the CR propagation model adopted in our analysis.
In Section \ref{section_cr_injection_sources_}, we outline the injection spectra of $e^{\pm}$ for the background, DM annihilation, and single nearby pulsar.
%In Section \ref{section_goodness_of_fit_test}, we introduce the goodness of fit test adopted in this work.
%In Section \ref{section_dm_fit_ams-02_leptonic_data_}, we fit to the AMS-02 lepton data and assess the quality of fit.
In Section \ref{section_can_local_known_single_pulsar_confuse_dampe_to_confirm_dm_signal}, we use the DAMPE data to investigate the DM annihilation and the single pulsar explanations for high energy $e^\pm$.
Finally, the conclusion is given in Section \ref{section_conclusion}.

\section{CR $e^\pm$ propagation in the Galaxy} \label{section_electron_(positron)_propagation_in_galaxy}

Galactic supernova remnants (SNRs) are generally believed to be the main source of primary CR particles with energies below $\sim 10^{17}\, \ev$.
After leaving the source, CR particles travel along the trajectories which are tangled by the Galactic magnetic field, and thus diffusively propagate.
Furthermore, they would also suffer from the so-called re-acceleration effect by scattering with the moving magnetic turbulence and gaining energy through the second order Fermi acceleration.

CRs propagate within a magnetic cylindrical diffusion halo with a characteristic radius of $20\, \kpc$ and a half height $z_h \sim \mathcal{O}(1)$ kpc. At the boundary of the propagation halo, CRs would freely escape.
During the journey to the earth, CRs lose their energies by a variety of effects; the primary nucleons fragment through inelastic collisions with the interstellar medium (ISM) and create secondary CR particles.

\begin{savenotes}
\begin{table}[!htbp]
		\caption{The mean values and $1\sigma$ uncertainties of the propagation and proton injection parameters for the DR2 propogation model. } 	\label{tab:pro}
	\centering
	\begin{tabular}{lcl}\hline \hline

		$D_0$		&$10^{28} {\cm}^2 \sec^{-1}$ &$4.16 \pm 0.57$ \\
		$\delta$ &						 		&$0.500 \pm 0.012$\\
		$z_h$		&$\kpc$						&$5.02 \pm 0.86$\\
		$v_A$		&$\km \, \sec^{-1}$			&$18.4 \pm 2.0$\\
		$R_0$		&$\GV$						& 4\\
		$\eta$	&								&$-1.28 \pm 0.22$\\
		$\mathrm{log(A_{\rho})}$\footnote{Propagated flux normalization at $100\, \GeV$ in unit of $\cm^{-2}\sec^{-1}\sr^{-1} {\MeV}^{-1}$}
					&					&$-8.334 \pm 0.002$\\
		$\nu_1$	&								&$2.04 \pm 0.03$\\
		$\nu_2$	&								&$2.33 \pm 0.01$\\
		$\mathrm{log(R_{br}^p)}$\footnote{Break rigidity of proton injection spectrum in unit of MV}
					&					&$4.03 \pm 0.03$\\
\hline
	\end{tabular}
\end{table}
\end{savenotes}

Involving the diffusion, re-acceleration, momentum loss and fragmentation effects, the transport equation can be described as \cite{ginzburg1964origin,ginzburg1990astrophysics}

\begin{equation}\label{eq:transp}
	\frac{\partial \Psi(\vec{r},p,t)}{\partial t} = Q(\vec r,p) + \nabla\cdot(D_{xx}\nabla\Psi) + \frac{\partial}
	{\partial p}p^2 D_{pp} \frac{\partial}{\partial p} \frac{1}{p^2}\Psi -\frac{\partial}{p}
	\dot{p} \Psi -\frac{\Psi}{\tau_f} ,
\end{equation}
where $\Psi(\vec{r},p,t)$ is the CR density per unit momentum interval at $\vec r$, $Q(\vec r,p)$ is the source term including primary and spallation contributions,
$D_{xx}$ is the spatial diffusion coefficient, $D_{pp}$ is the diffusion coefficient in momentum space,
$\tau_f$ is the time scale for the loss by fragmentation.
We use the public code GALPROP \cite{strong1999galprop,strong2009galprop,strong2015recent}\footnote{\url{http:galprop.stanford.edu/}} to numerically solve  this equation.

The spatial diffusion coefficient is described by

\begin{equation}
	D_{xx} = {\beta}^{\eta} D_0 {(\frac{R}{R_0})}^{\delta} ,
\end{equation}
where $\beta = {\upsilon}/{c}$ is velocity of particle in unit of the speed of light,
$D_0$ is a normalization constant, $R = {pc}/{ze} $ is the rigidity, $R_0$ is the reference rigidity and
$\eta$ describes the velocity dependence of the diffusion coefficient.

The re-acceleration effect can be described by the diffusion in the momentum space. The momentum diffusion coefficient $D_{pp}$ and spatial diffusion coefficient $D_{xx}$ are related by \cite{seo1994stochastic}
\begin{equation}\label{eq:re}
    D_{pp} = \frac{1}{D_{xx}} \cdot \frac{4p^2 {V_A}^2 }{3 \delta(4-\delta)(4-{\delta}^2)\omega},
\end{equation}
where $\omega$ denotes the level of the interstellar turbulence.
Absorbing $\omega$ to $V_A$ and referring $V_A$ characterizes the re-acceleration strength.

In Ref. \cite{Yuan:2017ozr} we have systematically studied the typical propagation models and nuclei injection spectra using the latest Boron-to-Carbon ratio $\mathrm{B/C}$ data from AMS-02 and the proton fluxes from PAMELA and AMS-02.
We find that the DR2 model including the re-acceleration and velocity depending diffusion effects gives the best fit to all the data. The posterior mean values and $68\%$ confidence interval of the model parameters are given in table \ref{tab:pro}.

The local interstellar flux is given by $\Phi = \Psi(r_{\odot})c/4\pi$.
Before the local interstellar (LIS) CRs arrive at the earth, they suffer from the solar modulation effect within the heliosphere. We employ the force field approximation, which is described by a solar modulation potential $\phi$,
to deal with this effect.

\section{CR injection sources } \label{section_cr_injection_sources_}

The observed CR $e^\pm$ consist of three components:
the primary electrons produced by SNRs;
the secondary electrons and positrons from primary nuclei spallation processes in the ISM;
$e^\pm$ pairs generated from exotic sources such as the DM or pulsar.
The sum of the first two components is treated as the background.
In this Section, we outline the injection CR $e^{\pm}$ spectra for the backgrounds, DM annihilations, and single nearby pulsar.

\subsection{The $e^{\pm}$ background spectrum}

Ordinary CR sources are expected to be located
around the Galactic disk, following the SNR radial distribution given by \cite{trotta2011constraints}
\begin{equation}\label{eq:frz}
	f(r,z) = (r/r_{\odot})^{1.25} \exp(-3.56 \cdot  \frac{r-r_{\odot}}{r_{\odot}}) \exp(-\frac{|z|}{z_s}) ,
\end{equation}
where $r_{\odot} = 8.3\, \kpc$ is distance between the sun and the Galactic center, $z_s = 0.2 \, \kpc$ is the characteristic height of the Galactic disk.

These sources are able to accelerate the high energy CR electrons through the first order Fermi shock acceleration, which would result in a power law spectrum.
The previous studies have found that a three-piece broken is enough to describe the injection spectrum of electrons below $\sim\TeV$\cite{Moskalenko:1997gh,lin2015quantitative}.
The break at a few $\GeV$ is used to fit the low energy data, while the hardening around hundreds of $\GeV$ is introduced to account for the effect of possible nearby sources~\cite{Bernard:2012pia} or non-linear particle acceleration~\cite{Ptuskin:2012qu}.

Above $\TeV$, the contribution to observed CR electrons would be dominated by several nearby SNRs due to the serious energy loss effect. The electron spectra from these SNRs may depend on their properties.
The detailed discussions can be found in Ref.~\cite{fang2016perspective,Fang:2017tvj}.
In this work, we simply introduce an exponential cutoff $\sim\TeV$ to describe the behaviour of high energy injection spectra from the nearby SNRs.
Thus, the injection spectrum of the primary electron component follows the form
\begin{equation}\label{eq:prminj}
	q(R) \propto 	\left\{
			\begin{array}{c}
                (R/R_{br})^{-\gamma_0}\exp(E/E_{bgc})  ,\quad R<=R_{br0} \\
                (R/R_{br})^{-\gamma_1}\exp(E/E_{bgc})  ,\quad R<=R_{br} \\
                (R/R_{br})^{-\gamma_2}\exp(E/E_{bgc})  ,\quad R>R_{br}
	        \end{array}
	\right.
  .
\end{equation}
The source function $Q(\vec r ,p)$ in Eq.~\ref{eq:transp} for SNRs is given by $Q(\vec r,p) = f(r,z)\cdot q(R)$.

The secondary $e^\pm$ are produced by the spallation of primary particles (mainly protons and helium nuclei) in the ISM.
The steady-state production rate of the secondary $e^\pm$ at the position $\vec r$ is
\begin{equation}\label{eq:secinj}
  Q_{sec}(\vec r,E) = 4\pi \sum_{ij} \int \mathrm{d}E^{'} \Phi_i(E^{'},\vec r) \frac{\mathrm{d}\sigma_{ij}(E^{'},E)}{\mathrm{d}E} n_j(\vec r),
\end{equation}
where $\mathrm{d}\sigma_{ij}(E',E)/\mathrm{d}E$ is the differential cross section for $e^\pm$ with the kinetic energy of $E$ from the interaction between the CR particle $i$ with the energy of $E'$ and ISM target $j$, and $n_j$ is the number density of the ISM target $j$.

\subsection{$e^{\pm}$ from DM annihilations}

The Galaxy is embedded in a huge DM halo.
If DM particles have some interactions with standard model particles, DM annihilations could produce CRs as an exotic source.
The source term for DM annihilations is given by
\begin{equation}\label{eq:dmsource}
	Q_{DM}(\vec r,E) = \frac{1}{2} {\frac{ \rho_{DM}(\vec r)}{m_{DM}}}^2 \langle {\sigma \upsilon}\rangle \sum_k{ B_k \frac{dN^k_{e^\pm}}{dE}} ,
\end{equation}
where $\mathrm{d}N^k_{e\pm}/\mathrm{d}E$ denotes the $e^{\pm}$ energy spectrum from a single annihilation with final states $k$, $B_k$ is the corresponding branching fraction, $\rho$ is the DM density, and $\langle {\sigma \upsilon}\rangle$ is thermal averaged velocity-weighted annihilation cross section.
In our analysis, we adopt the Navarro-Frenk-White (NFW) density profile \cite{navarro1997universal}
\begin{equation}\label{eq:profile}
	\rho_{DM}(\vec r) = \frac{\rho_s}{(|\vec r|/r_s){(1+ |\vec r|/r_s)}^2},
\end{equation}
where $r_s = 20 \,\kpc$, and the local DM density is normalized to $\rho_{\odot}= 0.4 \,\GeV\cm^{-3}$ consistent with the dynamical constraints \cite{nesti2013dark,sofue2012grand,weber2010determination}.
The initial energy spectra of DM annihilations are taken from PPPC 4 DM ID \cite{cirelli2011pppc} which includes the electroweak corrections.
We also use GALPROP to simulate the propagation of such emissions from DM annihilation.

\subsection{$e^{\pm}$ from the nearby pulsar}\label{sec:pssoc}

The pulsar is a rotating neutron star surrounded by the strong magnetic field. It can produce $e^{\pm}$ pairs through the electromagnetic cascade and accelerate them by costing the spin-down energy.
These high energy $e^{\pm}$ pairs are injected into the pulsar wind nebula (PWN) and finally escape to the ISM.
A burst-like spectrum of the electron and positron is adopted to describe the pulsar injection, and is usually assumed to be a power law with an exponential cutoff
\begin{equation} \label{eq:pulsarinj}
	Q_{psr}(\vec r,E,t) = Q_0 E^{-\alpha} \exp{(-\frac{E}{E_c})} \delta(\vec r) \delta(t),
\end{equation}
where $Q_0$ is the normalization factor, $\alpha$ is the spectrum index, and $E_c$ is the cutoff energy.
The total energy output is related to the spin-down energy $W_0$ by assuming that a fraction $f$ of $W_0$ would be transferred to $e^{\pm}$ pairs, so that
\begin{equation}
    \int_{0.1\GeV}^{\infty} \mathrm{d}E E Q_0 E^{-\alpha} \exp{(-\frac{E}{E_c})} = \frac{f}{2} W_0.
\end{equation}
$W_0$ can be derived from $\tau_0 \dot{E}(1+t/{\tau_0})^2$, where $t$ is the pulsar age and the typical pulsar decay time is taken to be $t_0 = 10^{4}\, \kyr$ here \cite{aharonian1995high,profumo2012dissecting}.

Due to the high energy loss rate, the energetic $e^{\pm}$ could only propagates over a small distance of $\mathcal{O}(1)$ kpc \cite{yin2013pulsar,feng2016pulsar}.
For the local sources, we adopt an analytic solution of the propagation equation.
The re-acceleration effect is neglected for particles with energies above $10 \,\GeV$ \cite{delahaye2009galactic}.
The Green function solution without boundary condition is given by \cite{delahaye2010galactic,fang2016perspective}
\begin{equation} \label{eq:locflux}
	\Psi(r_{\odot},E) = \frac{1}{(\pi \lambda^2(E,E_s))^{3/2}}\cdot \exp{(-\frac{r^2}{\lambda^2(E,E_s)})} \cdot Q_{psr}(E_s),
\end{equation}
where $r$ is the distance to the pulsar, $E_s$ is the initial $e^{\pm}$ energy from the source, and $\lambda$ is the diffusion length defined as
\begin{equation}
\lambda^2 \equiv 4 \int_E^{E_s} \mathrm{d}E^{'} {D(E')}/{b(E^{'})}.
\end{equation}
The $E_s$ could be derived from the propagation time $t$ and the final energy $E$ by the relation of
\begin{equation} \label{eq:Es}
  \int_{E_s}^E - \frac{\mathrm{d} E}{b(E)} = \int_{-t}^0\mathrm{d}t,
\end{equation}
where $b(E)\equiv -\mathrm{d}E/\mathrm{d}t$ is the energy loss rate of $e^{\pm}$.
We include the energy loss induced by the synchrotron radiation in the Galactic magnetic field and the inverse Compton scattering with the ambient photon field.
The local interstellar radiation field (ISRF) is taken from M1 model in Ref. \cite{delahaye2010galactic} and the magnetic field is assumed to be $4 \,\mu G$.
A detailed discussion on the relativistic energy loss rate is given in appendix \ref{sec:lossrate}.
As no simple analytic solution is available, we use GNU Scientific Library (GSL) \footnote{\url{https://www.gnu.org/software/gsl/}} to numerically solve Eq.\ref{eq:Es}.

\section{Explanations of the DAMPE $e^\pm$ spectrum} \label{section_can_local_known_single_pulsar_confuse_dampe_to_confirm_dm_signal}

The DAMPE collaboration has reported the $e^\pm$ spectrum with high resolution from $25 \GeV$ to $4.6 \TeV$~\cite{Ambrosi:2017wek}.
In this section, we fit the DAMPE data in the DM annihilation and single pulsar scenarios.
The AMS-02 positron fraction ~\cite{accardo2014high} is also considered in the fit.
Since we focus on the high energy $e^\pm$, the positron fraction data below 10 $\GeV$ are not adopted.
There are 81 data points involved in the analysis.

\subsection{The DM scenario} \label{subsection_dm_scenarios}
DM particles in the propagation halo may decay or annihilate to standard model particles and contribute to the finally observed CR leptons \cite{Bergstrom:2000pn,Bertone:2004pz,Bergstrom:2009ib}.
In this work, we discuss the DM annihilation channels to $\mu^+\mu^-$, $\tau^+\tau^-$, $e\mu\tau$, $4\mu$, and $4\tau$.
In the $e\mu\tau$ channel, we set the branch ratios of $e^+e^-$, $\mu^+\mu^-$ and $\tau^+\tau^-$ final states to be free parameters.
With the propagation parameters and proton injections given in Table.~\ref{tab:pro}, we vary the injection of primary electrons, DM parameters and solar modulation potential to obtain the best-fit though the Monte Carlo Markov Chain (MCMC) method.
Note that we only consider the energy region above $10\GeV$, thus the low energy break around several $\GeV$ in the background electron spectrum is neglected.
In addition, a rescale factor $c_{e^\pm}$ is introduced to indicate the uncertainty of the hadronic collisions.
The degree of freedom (d.o.f.) in this fitting is thus 72.\footnote{There is also a normalize factor of the injection spectrum $A_e$, which is not quite relevant but is still taken as a free parameter.}

\begin{table}[!htbp]
     \caption{Posterior mean and $68\%$ credible uncertainties of the model parameters and $\chi^2$ value in the DM scenarios, with d.o.f. of 72.}     \label{tab:dmfit}
     \centering
 \begin{tabular}{c c c c c c}
     \hline

                   &$\mu^{+}\mu^{-}$   &$\tau^{+}\tau^{-}$ &$e\mu\tau$    &$4\mu$       &$4\tau$ \\\hline
     $\gamma_1$    &$3.05 \pm 0.02$      &$2.93 \pm 0.02$        &$2.93 \pm 0.03$     &$3.06 \pm 0.01$   &$2.90 \pm 0.02$   \\
     $\gamma_2$    &$2.57 \pm 0.02$      &$2.50 \pm 0.02$        &$2.53 \pm 0.02$     &$2.53 \pm 0.02$   &$2.50 \pm 0.02$      \\
     $\log(R_{br}^e/\mathrm{MV})$ &$4.69\pm0.03$      &$4.70\pm0.04$  &$4.73\pm0.03$  &$4.71 \pm 0.03$   &$4.68 \pm 0.04$      \\
     $c_{e^{\pm}}$ &$3.66 \pm 0.03$      &$2.82 \pm 0.09$        &$3.07 \pm 0.19$     &$3.67 \pm 0.04$   &$2.65 \pm 0.03$ \\
     $E_{bgc}/\TeV$     &$4.42 \pm 0.96$       &$6.2 \pm 2.0$        &$10.9\pm2.2$      &$4.17 \pm 1.09$   &$5.58 \pm 0.71$ \\
     $\phi/\mathrm{GV}$ &$1.48\pm0.02$   &$0.77 \pm 0.10$        &$1.16 \pm 0.17$     &$1.30 \pm 0.08$   &$0.67\pm  0.07$ \\\hline
     $m_{DM}/\mathrm{GeV}$        &$1891\pm71$        &$3210\pm316$   &$1560\pm178$   &$3243 \pm 290$    &$5366 \pm 338$   \\
     $\langle\sigma\nu\rangle/(10^{-23}{\mathrm{cm}}^3\sec^{-1})$  &$1.37\pm0.09$ &$5.24\pm0.82$ &$1.46\pm0.31$ &$2.25\pm0.37$ &$7.96\pm1.01$ \\\hline
     $\chi^2$      &113.30               &68.71                  &59.69               &89.87             &69.56  \\
\hline
 \end{tabular}
 \end{table}

\begin{figure}[!hbt]
  \centering
  \includegraphics[width=0.45\textwidth]{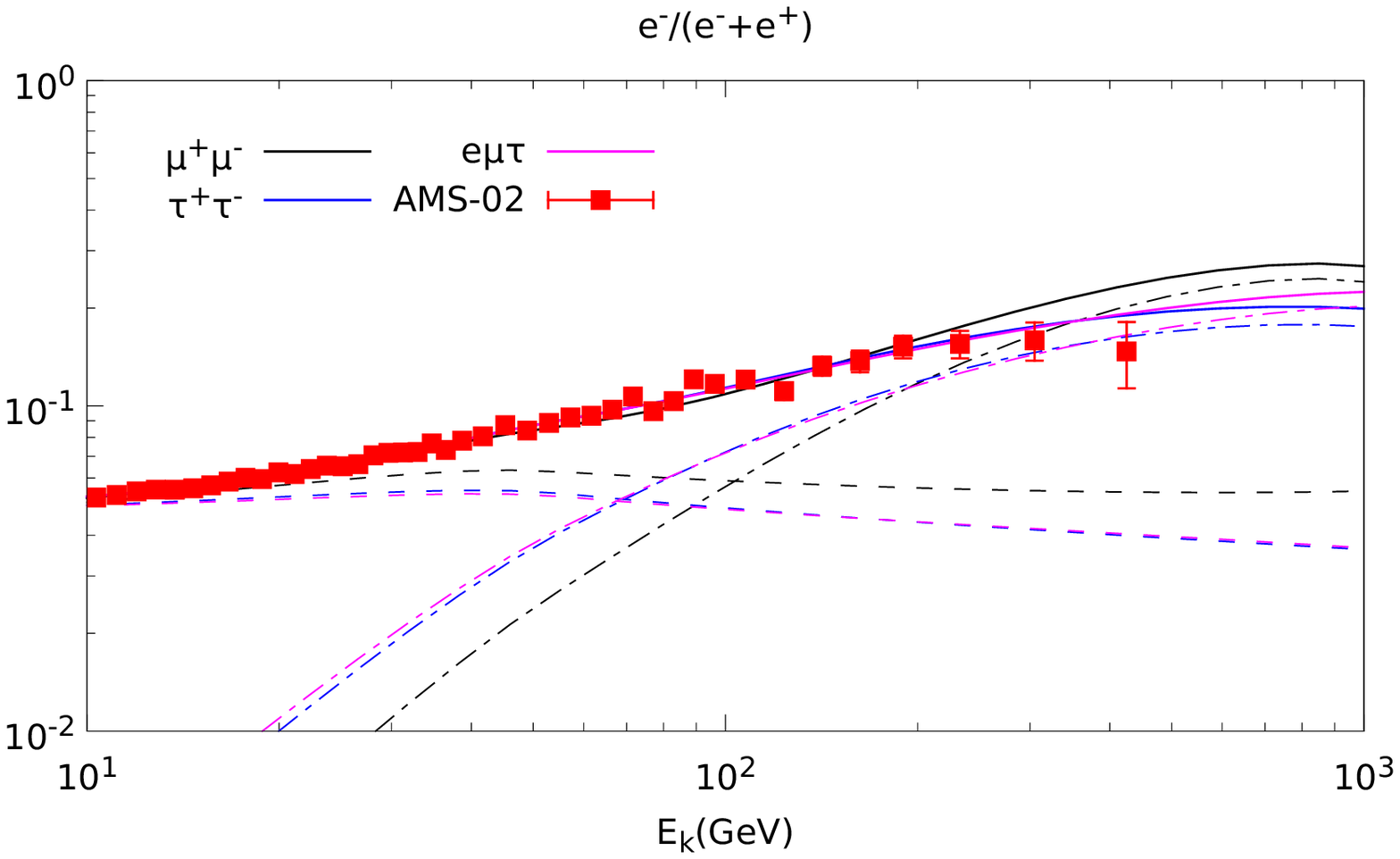}
  \includegraphics[width=0.45\textwidth]{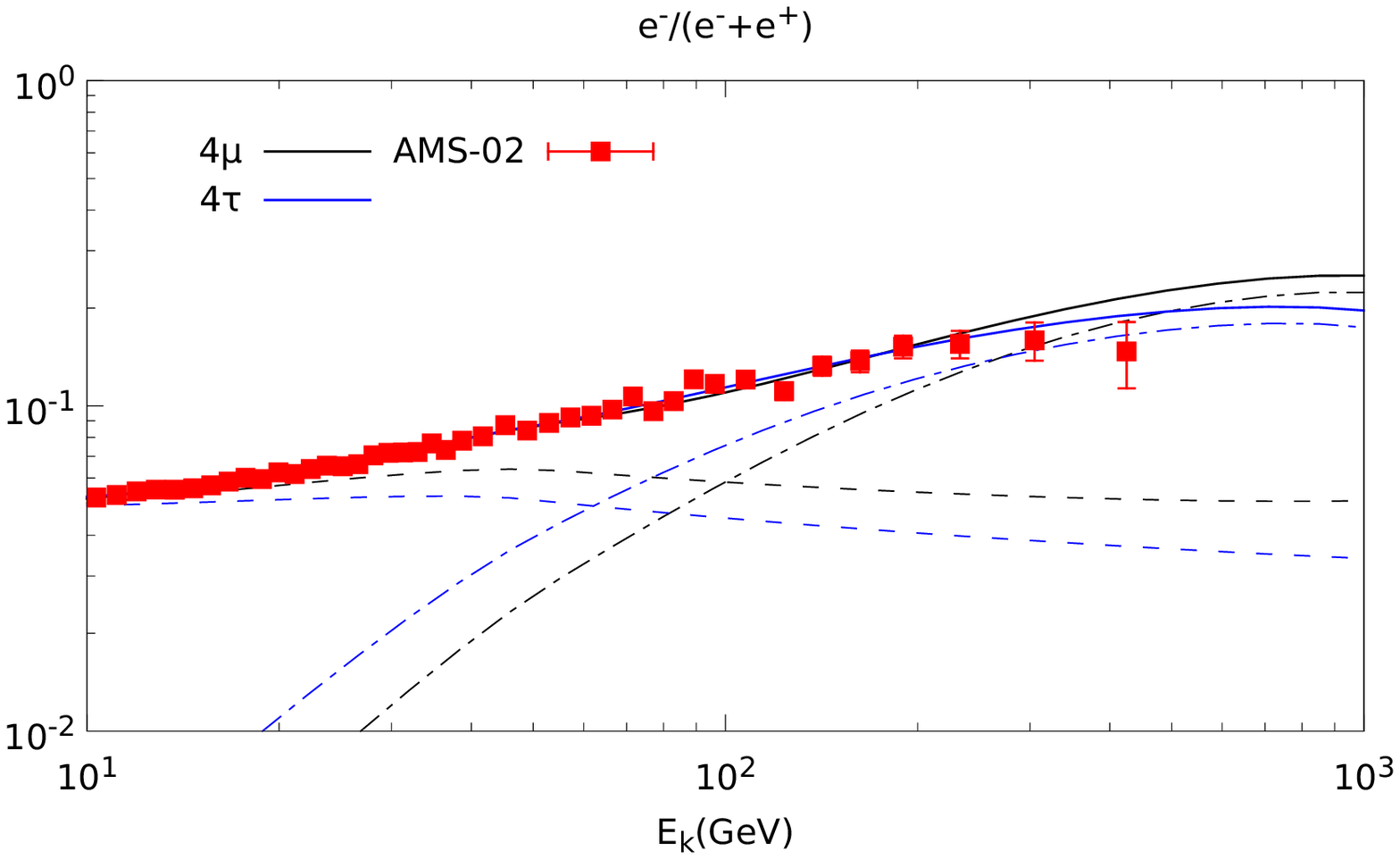}\\
  \includegraphics[width=0.45\textwidth]{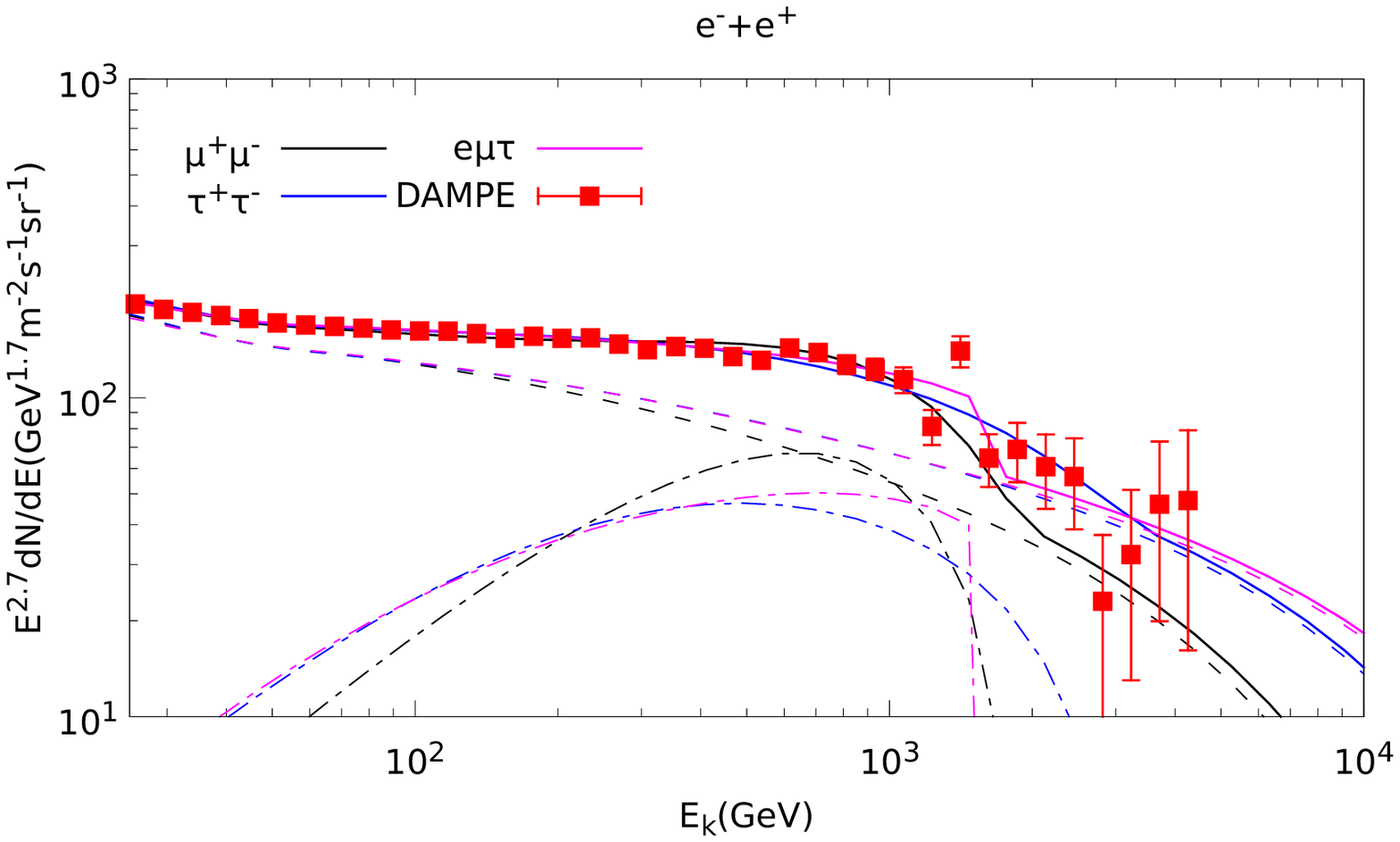}
  \includegraphics[width=0.45\textwidth]{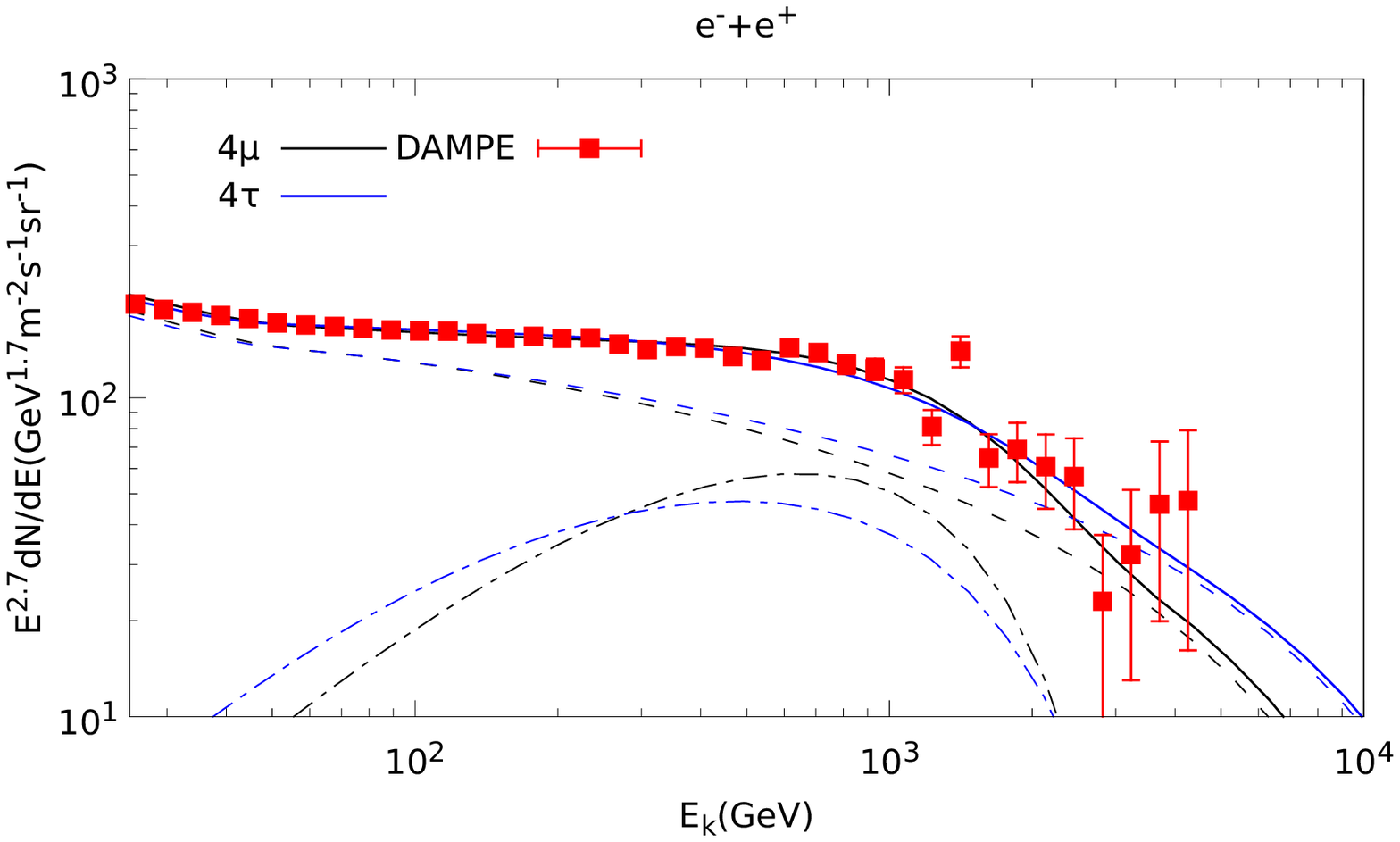}
  \caption{The spectra of the best-fit restuffs in the DM scenario. Top panels show the positron fractions in comparison with the AMS-02 data~\cite{accardo2014high}, while the bottom panels show the total $e^\pm$ spectra in comparison with the DAMPE data~\cite{Ambrosi:2017wek}. The dashed, dotted-dashed and solid lines represent the backgrounds, DM contributions and total results, respectively.}
  \label{fig:dm_best}
\end{figure}

We list the best-fit results in Table.~\ref{tab:dmfit} and show the spectra in Fig.~\ref{fig:dm_best}.
With a large $\chi^2$ valued 113.3, the $\mu^+\mu^-$ channel is excluded with more than $3\sigma$ confidence, while all the other channels provide reasonable fits.
This is because that the $\mu^+\mu^-$ channel induces a harder $e^\pm$ spectrum than all the other channels, and tends to produce too many positrons at high energies when explaining the positron fraction below $100\GeV$.

Note that the fit for annihilation channels with hard DM contributions is sensitive to the secondary positrons.
In Ref.~\cite{Yuan:2017ysv}, Yuan et al. performed a similar analysis but with a smaller diffusion coefficient power index $\delta=1/3$.
Since this $\delta$ leads to a harder secondary positron spectrum, the $\mu^+\mu^-$ channel work well in that analysis.

In the fit for the $e\mu\tau$ channel, since the sharp shapes of the injection spectra from the $\mu^\pm$ and $e^\pm$ final states are not favoured here, the $\tau^\pm$ final states are dominant with a branch ratio of 0.755, while he branch ratios of $e^\pm$ and $\mu^\pm$ are 0.094 and 0.151 respectively .
A recent work also analysed this channel but found that the branch ratio of $\tau^\pm$ is suppressed~\cite{Niu:2017hqe}.
This different conclusion may be attributed by that their background secondary positron spectrum is much harder than ours.
It is interesting to note that the contribution from the $e^\pm$ final states would indicate a distinct drop in the spectrum at the DM mass, as shown in the bottom left panel of Fig.~\ref{fig:dm_best}.
It is possible to check such spectral feature in the results of DAMPE or HERD in the future.

\begin{figure}[!hbt]
  \centering
  \includegraphics[width=0.7\textwidth]{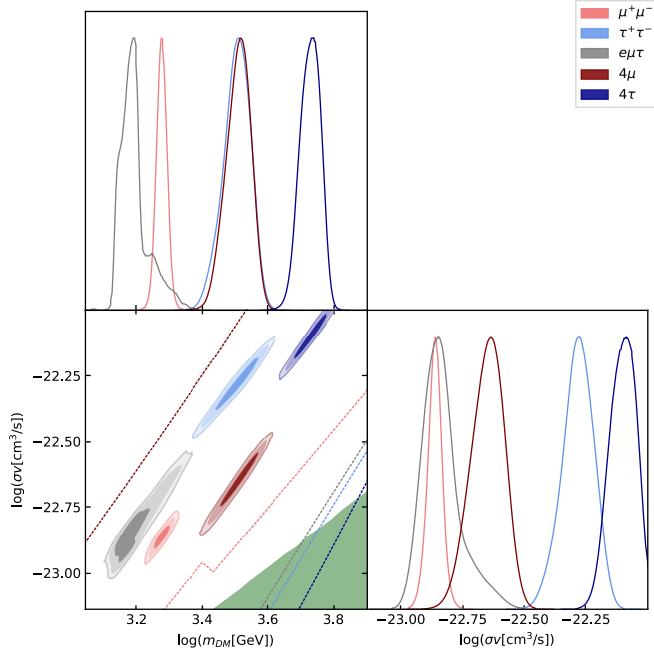}
  \caption{The $68\%$ and $95\%$ confidence regions of the DM mass and thermally averaged annihilation cross section for different annihilation channels. The corresponding limits derived from the Fermi-LAR gamma-ray observations of dwarf galaxies~\cite{Fermi-LAT:2016uux} are also shown. The dark green shade region represents the constraint from the CMB observations, with an energy deposition efficiency $f_\mathrm{eff}$ ranged from 0.15 to 0.7~\cite{Ade:2015xua}.}
  \label{fig:dm_contour}
\end{figure}

In Fig.~\ref{fig:dm_contour}, we show the $68\%$ and $95\%$ confidence regions for the DM mass and thermally averaged annihilation cross section.
Note that the DM implication for the CR $e^\pm$ excess has been strongly constrained by many other observations, such as the
cosmic microwave background (CMB)~\cite{komatsu2009five,ade2016planck,slatyer2016indirect}, dwarf galaxy gamma-ray ~\cite{collaboration2015searching}, and diffuse gamma-ray observations \cite{ackermann2012constraints}.
We also show the corresponding constraints in Fig.~\ref{fig:dm_contour}.
To calculate the constraints from dwarf galaxies, we use PPPC 4 DM ID \cite{Cirelli:2010xx} to produce the DM gamma-ray spectra and adopt the likelihood results from the combined analysis given in ~\cite{Fermi-LAT:2016uux}.
The constraints from the CMB observations are taken from Ref.~\cite{Yuan:2017ysv}, where the Planck 2015 results with an energy deposition efficiency $f_\mathrm{eff}$ ranged from 0.15 to 0.7~\cite{Ade:2015xua} are adopted.
The best-fit regions for many channels shown in Fig.~\ref{fig:dm_contour} seem to be excluded.
However, the tensions between these observations can be reconciled in the velocity-dependent annihilation scenario due to the fact that DM particles contributing to these observations have different typical relative velocities~\cite{Bai:2017fav,xiang2017dark}.

On the other hand, the constraints from the diffuse gamma-ray observation \cite{ackermann2012constraints} cannot be easily avoided in the velocity-dependent annihilation scenario.
However, the astrophysical uncertainties of this analysis arising from the Galactic CR model are not negligible.
The solid constraints given by the Fermi-LAT collaboration without modeling of the astrophysical background cannot exclude the DM implication accounting for the CR $e^\pm$ excess.
When the contributions of the pion decay and inverse Compton scattering from Galactic CRs are considered, the constraints given by the Fermi-LAT collaboration become stringent and strongly disfavor the $\tau^+\tau^-$ channel.
However, these constraints depend on the CR distribution and prorogation models.
In order to reduce the related uncertainties, more precise data and further studies on the CR model will be needed.

\subsection{The single pulsar scenario}

The nearby pulsars are possible sources of high energy CR positrons.
We consider the cases in which a nearby mature pulsar is the primary source of high energy $e^\pm$.
The 21 known pulsars within $1\, \kpc$ with characteristic ages in the range of $10^3-10^6$ years are considered.
Their properties, such as the distance, age and spin-down luminosity, are taken from the ATNF catalog~\cite{manchester2005australia}, which includes the most exhaustive and updated list of known pulsars.

In this section, we fit the DAMPE total $e^\pm$ spectrum and the AMS-02 positron fraction using one of the 21 selected pulsars.
Comparing with the DM scenario with two free parameters, there are three free parameters ($f$, $\alpha$, $E_c$) in the pulsar scenario.
Thus the d.o.f. is 71 in the fit.
Then, we drop all the candidates which are not acceptable at 95.4\% C.L., and list the best-fit result of the five left pulsars in Table.~\ref{tab:psfit}.

\begin{table}[!htbp]
  \caption{Posterior mean and $68\%$ credible uncertainties of the model parameters and $\chi^2$ value in the single pulsar scenarios, with d.o.f. of 71.}
  \label{tab:psfit}
	\centering
	%\resizebox{\columnwidth}{!}{
  \begin{tabular}{cccccc}
\hline
                     &     J1732-3131   &   J0940-5428   &    B0656+14    &    B1001-47    & J0954-5430   \\
                     &                  &                &    (Monogem)   &                &     \\\hline
     $\gamma_1$      &   $2.94\pm0.03$  &$2.93\pm0.04$   &$2.84\pm0.03$   &$2.80\pm0.03$   &$2.84\pm0.03$        \\
     $\gamma_2$      &   $2.52\pm0.02$  &$2.51\pm0.03$   &$2.52\pm0.03$   &$2.50\pm0.02$   &$2.52\pm0.02$      \\
$\log(R_{br}^e/\MV)$ &   $4.72\pm0.05$  &$4.73\pm0.05$   &$4.81\pm0.05$   &$4.84\pm0.04$   &$4.81\pm0.05$      \\
   $E_{bgc}/\TeV$    &   $6.41\pm2.25$  &$4.99\pm2.67$   &$5.93\pm3.02$   &$7.91\pm1.56$   &$6.82\pm1.99$      \\
   $c_{e^{\pm}}$     &   $3.21\pm0.22$  &$3.18\pm0.24$   &$2.66\pm0.13$   &$2.37\pm0.20$   &$2.66\pm0.13$      \\
     $\phi/\GV$      &   $1.18\pm0.22$  &$1.17\pm0.25$   &$1.32\pm0.14$   &$1.18\pm0.21$   &$1.24\pm0.18$      \\\hline
        $f$          &   $3.59\pm0.88$  &$0.26\pm0.07$   &$2.86\pm0.25$   &$3.33\pm0.26$   &$7.58\pm0.62$      \\
      $\alpha$       &   $2.14\pm0.06$  &$2.08\pm0.07$   &$1.78\pm0.05$   &$1.82\pm0.04$   &$1.85\pm0.04$      \\
     $E_c/\TeV$      &   $4.94\pm2.12$  &$2.12\pm0.80$   &$2.65\pm0.93$   &$6.50\pm2.35$   &$5.46\pm1.89$      \\\hline
       $d/\kpc$      &        0.64      &      0.38      &      0.29      &      0.37      &  0.43    \\
 $\mathrm{age}/\kyr$ &        111       &      42.2      &      111       &      220       &  171   \\\hline
      $\chi^2$       &        59.82     &      62.98     &      56.13     &      60.64     &  56.39     \\ \hline
\end{tabular}

\end{table}

\begin{figure}[!hbt]
    \centering
	\includegraphics[width=0.45\textwidth]{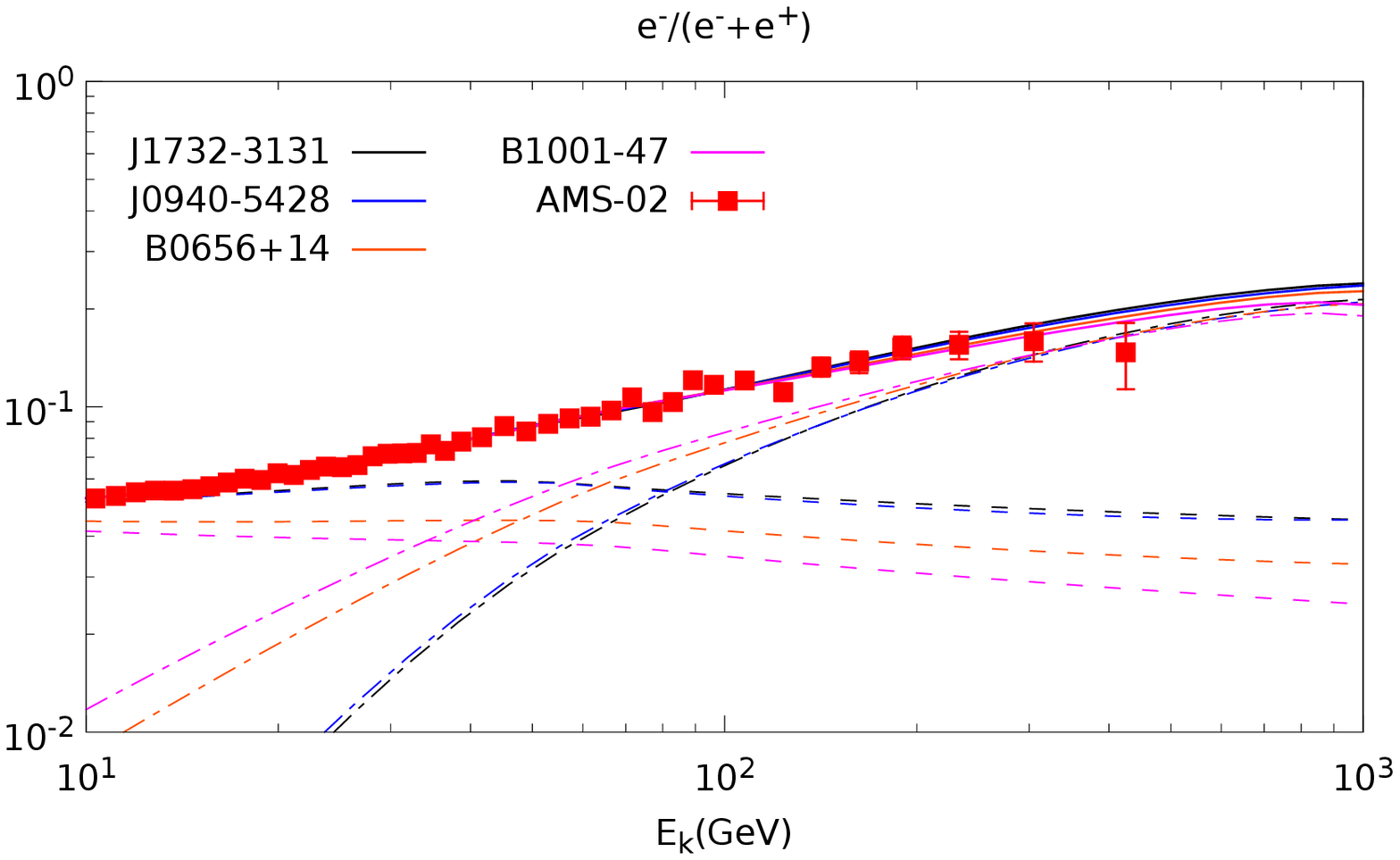}
	\includegraphics[width=0.45\textwidth]{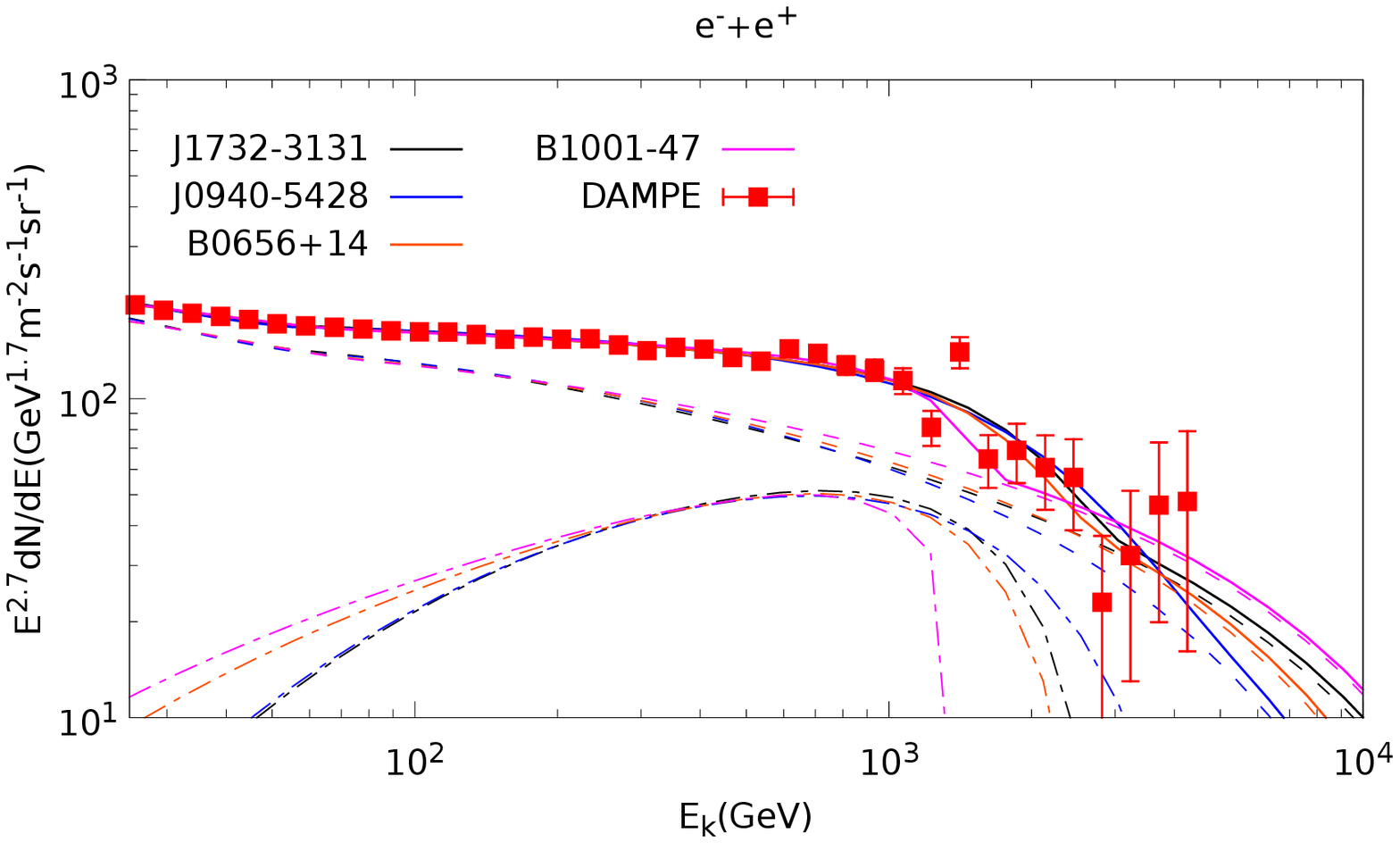}\\
  \caption{The same as Fig.~\ref{fig:dm_best}, but for the single pulsar scenario.}
    \label{fg:psfit_best}
\end{figure}

Note that there are uncertainties in the estimation for the spin-down energies of pulsars.
Therefore, a transfer fraction $f$ larger than 1 within a tolerance range may be allowed.
Except for J0954-5430 with a very large transfer fraction is 7.58, we accept the other four pulsars listed in Table.~\ref{tab:psfit}.
For these four acceptable pulsars, we show the best-fit spectra in Fig.~\ref{fg:psfit_best}.
We find that B1001-47, which is older than the others, would lead to a drop around $1.2\TeV$ at the $e^\pm$ spectrum, due to the energy loss effect. From Fig.~\ref{fig:dm_best} and Fig.~\ref{fg:psfit_best}, it can be seen that the best-fit spectra resulted from some pulsars are very similar to those induced by DM annihilation. It is still difficult to distinguish between these two explanations of high energy $e^\pm$ with the current DAMPE result.

\begin{figure}[!htp]
  \centering
  \includegraphics[width=0.8\textwidth]{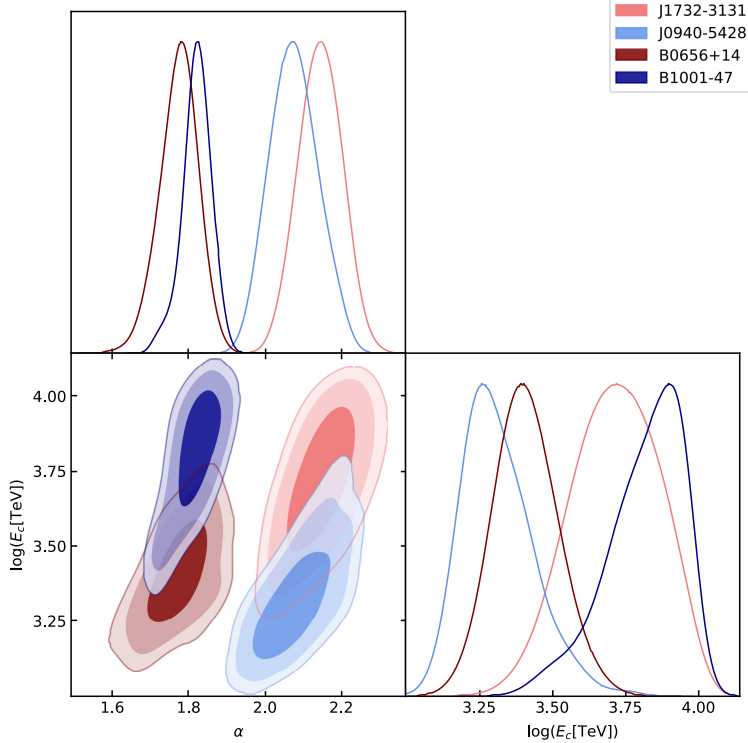}
  \caption{Similar to Fig.~\ref{fig:dm_contour}, but for the $\alpha$ and $E_c$ in single pulsar scenario.}
  \label{fig:ps_contour}
\end{figure}

In order to take account the uncertainty from the injection spectrum of the pulsar, the 68\% and 95\% confidence regions for the power index $\alpha$ and cutoff energy $E_c$ are shown in Fig.~\ref{fig:ps_contour}.
The favoured values of injection parameters depend on the age and distance of the pulsar.
For the far or young pulsars, such as J1732-3131 and J0940-5428, the injected low energy positrons are difficult to reach the Earth, therefore a large $\alpha$ is needed to result in enough positrons at low energies.
In addition, the contributions from such pulsars would suffer a serious energy loss effect, thus they require a high energy cutoff $E_c$ to ensure enough high energy positrons.

The local young astrophysical sources may induce the observable dipole anisotropy in the CR arrival direction \cite{shen1971anisotropy,kobayashi2004most,di2011implications,manconi2017dipole}.
Therefore, the four pulsars considered above would be constrained by the currently anisotropy observations.
Both Fermi and AMS-02 have not detected a significant anisotropy, and set upper limits on the CR electron-positron, positron and electron dipole anisotropies \cite{ackermann2010searches,abdollahi2017search,aguilar2013alpat,la2016search}.
For the AMS-02 experiment \cite{la2016search}, the $95\%$ C.L limit on the integrated positron dipole anisotropy based on the first five years data is about $0.02$ above $16 \,\GeV$.
Recently, the Fermi collaboration has released the latest result of the dipole $e^\pm$ anisotropy, using seven years data reconstructed by Pass 8 in the energy region from $42 \,\GeV$ to $2 \,\TeV$ \cite{abdollahi2017search}.
The $95\%$ C.L upper limit ranges from $3\times 10^{-3}$ to $3\times 10^{-2}$.

For the single nearby source dominating the CR flux, the dipole anisotropy is given by \cite{shen1971anisotropy}
\begin{equation}
	\Delta = \frac{3D}{c}|\frac{\nabla\Phi}{\Phi}| .
\end{equation}
By using the flux in Eq.~\ref{eq:locflux}, we explicitly express the dipole anisotropy as
\begin{equation}
	\Delta (E) = \frac{3D(E)}{c} \frac{2d_s}{\lambda^2(E,E_s)} .
\end{equation}

\begin{figure}[!hbt]\label{fg:ps_aniso}
	\centering
	\includegraphics[width=0.7\textwidth]{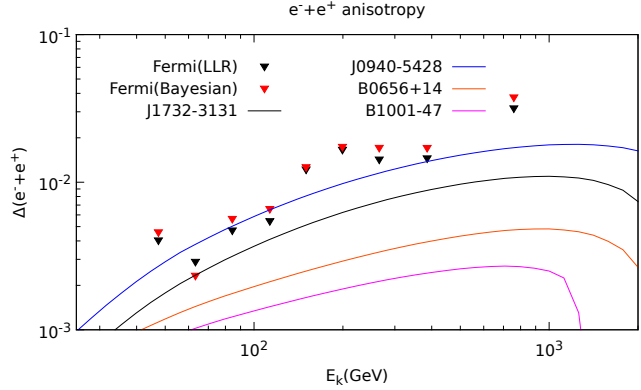}
  \caption{The $e^\pm$ anisotropies for the pulsars. The 95\% C.L. upper limits given by Fermi-LAT using the log-likelihood ratio and Bayesian methods \cite{abdollahi2017search} are also shown for comparison.}
\end{figure}

We estimate the positron anisotropy for the four pulsars, and find all of them would result in an integrated anisotropy obviously smaller than 0.02 at $16\GeV$, which evade the limit of AMS-02.
In addition, we show the expected $e^\pm$ anisotropy from each pulsar comparing with the measurement of Fermi-LAT in Fig.~\ref{fg:ps_aniso}.
Note that the dipole anisotropy is proportional to the distance-to-age ratio ${r}/{t}$ of the pulsar.
We find that the anisotropy induced by J0940-5428 with a ${r}/{t}\sim 9\times 10^{-3} $ kpc/kyr has a tension with the Fermi upper limit.
For J1732-3131 with a larger age of $111\,\kyr$ and a smaller ${r}/{t}\sim 5.8\times 10^{-3}$ kpc/kyr, there is a slight tension as the Fermi-LAT Bayesian limit excludes this source while the LLR limit does not.
The other pulsars with a very small ${r}/{t}\sim 2.5 \times 10^{-3}$ kpc/kyr provides a very small anisotropy, which is far from the Fermi-LAT sensitivity.
As a conclusion, B0656+14 and B1001-47 survive from all the limits considered.
%These results mean that the cases of DM annihilations to $\tau^+\tau^-$ and charged leptons are difficult to be distinguished from the single nearby pulsar case by DAMPE and current anisotropy results.

\section{Conclusion} \label{section_conclusion}

In this paper, we study the DM annihilation and pulsar interpretations of high energy CR $e^\pm$ observed by DAMPE.
We investigate the $e^\pm$ contributions from several DM annihilation channels and known nearby pulsars in the ATNF catalog, and find some allowed realizations.

For the DM scenario, we investigate the $\mu^+\mu^-$, $\tau^+\tau^-$, $e\mu\tau$, $4\mu$, and $4\tau$ annihilation channels.
We find that all these channels except for the $\mu^+\mu^-$ channel can explain the DAMPE data.
In the $e\mu\tau$ mixing channel, the $\tau^\pm$ final states are dominant.
However, the contribution from the $e^\pm$ final states in this channel would lead to a distinct drop in the spectrum.
Such spectral feature may be detected in the future measurements with larger statistics, such as DAMPE and HERD.
The constraints from the diffuse $\gamma$ ray, the $\gamma$ ray of dwarf galaxy and the CMB observations are discussed.
We find that many channels have been excluded and some complicated DM models are necessary to reconcile the tension between different observations.

For the pulsar scenario, we find five single pulsars that are acceptable at 94.5\% C.L.~.
Among these pulsars, J0954-5430 requires a too large transfer efficiency which is unacceptable.
We also investigate the $e^\pm$ anisotropy from the single pulsar, and find that J1732-3131 and J0940-5428 are not favored by the Fermi-LAT observations.
As a conclusion, B0656+14 and B1001-47 are possible to explain the current observations of DAMPE and AMS-02.
Our results show that it is difficult to distinguish between the DM annihilation and single pulsar explanations of high energy $e^\pm$ with the current DAMPE result.
In the future, the combination of the $e^\pm$ spectra and anisotropy measurements with large statistics like HERD may be useful to further discriminate the origins of high energy CR $e^\pm$.

\section*{Acknowledgment}
This work is supported by the National Key Program for Research and Development (No.
2016YFA0400200), by the 973 Program of China under Grant No.
2013CB837000, and by the National Natural Science Foundation of China
under Grants No. 11475189, 11475191,

\bibliography{paper}
\bibliographystyle{unsrt}

%\begin{appendices}
%\lipsum[1]
\newpage
\chapter{Appendix}
\appendix
%\lipsum[2-5]

\section{Energy loss and cutoff} \label{sec:lossrate}

By using Eq.~\ref{eq:Es} we can derive the $e^\pm$ energy $E_s$ in the source. The energy loss of $e^\pm$ above few $\GeV$ is mainly caused by the synchrotron radiation and inverse Compton scattering processes. The synchrotron radiation energy loss is given by
\begin{equation}\label{eq:synloss}
	-{(\frac{dE}{dt})}_{syn} = \frac{4}{3} \sigma_T c U_B {\gamma_e}^2 ,
\end{equation}
where $\sigma_T$ is the Thomson scattering cross section ($\sigma_T=6.65\times 10^{-25}\cm^2$), $c$ is the speed of light, $U_B$ is the magnetic field energy density, and $\gamma_e = E/{m_e c^2}$ is the Lorentz factor.

The energy loss caused by the inverse Compton scattering of $e^\pm$ with energies $E<<{(m_e c^2)}^2/{k_b T}$ in the Thomson regime can be expressed by Eq.~\ref{eq:synloss}) with replacing $U_B$ to the radiation field energy density ($U_{rad}$). The energy density distribution of ISRF is described in section \ref{sec:pssoc}.
In the Thomson approximation the energy loss rate can be given by $-\frac{dE}{dt} = b_0 E^2$, where $b_0 = \frac{4}{3 m_e c^2} \sigma_T c (U_B+U_{rad})$ is a constant. This number is often taken to be $\mathrm{O}(10^{-16}) \, \GeV^{-1}/{\mathrm{s}}$ \cite{blasi2009origin,feng2016pulsar}. Then we can derive the maximum energy of $e^\pm$ arriving at the solar system as
	\begin{equation}
		\int_{E_s}^E \frac{dE}{-b_0 E^2} = \int_{-t}^0 dt  \quad \rightarrow \quad E = \frac{1}{b_0 t + 1/E_s} < \frac{1}{b_0 t} .
	\end{equation}
This indicates that the maximum energies of observed $e^\pm$ are determined by $1/{b_0 t}$ due to the energy loss and are almost independent of their initial energies from the source.

However, under the extreme Klein-Nishina limit with $E>>{(m_e c^2)}^2/{k_b T}$, the energy loss rate is
	\begin{equation}
		-{(\frac{dE}{dt})}_{KN} = \frac{\sigma_T}{16} \frac{{(m_e c k_b T)}^2}{\hbar^3} {\mathrm{ln}\frac{4\gamma_e k_b T}{m_e c^2}  - 1.9805}.
	\end{equation}
In this case the energy loss rate only increases logarithmically with $E$, while it increases with $E^2$ under the Thomson limit. Thus the efficiency of inverse Compton scattering would be strongly reduced; this effect is referred to as "KLein-Nishina cutoff"\cite{schlickeiser2013cosmic} and has been discussed in Ref. \cite{kobayashi2004most,schlickeiser2010klein,delahaye2010galactic,khangulyan2014simple}. In this work we adopt the parametrization expression in \cite{delahaye2010galactic} to accurately calculate the energy loss rate caused by the inverse Compton scattering.

\begin{figure}[htbp]
     \includegraphics[width=1\textwidth]{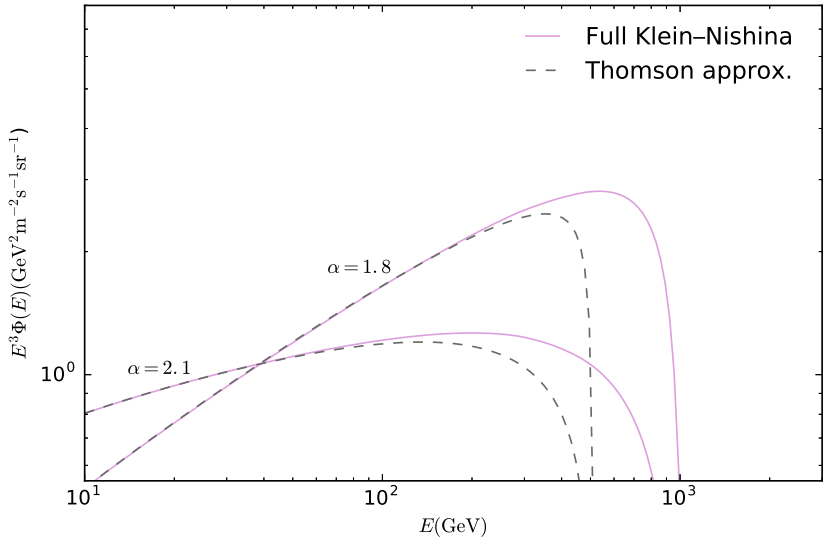}
	  \caption{Comparison between different treatments for the inverse Compton energy loss of the Geminga electron spectrum. The solid and dashed lines represented results for the relativistic energy loss rate and the Thomson approximation energy, respectively. $\alpha$ is the index of the injection spectrum.}
\label{fg:cutoff}
\end{figure}

For instance, we illustrate the Klein-Nishina effect for Geminga, which is a pulsar with a distance of $0.25 \, \kpc$, an age of $342 \, \kyr$, and a spin-down luminosity of $3.2 \times 10^{34} \, \erg/\sec$. The energy transfer efficiency $f$ is taken as $30\%$ and the injection spectrum energy cutoff is assumed to be $10 \, \TeV$. As can be seen in figure \ref{fg:cutoff}, since Thomson approximation results in a higher energy loss rate, the spectrum cutoff is sharper than that derived from the Klein-Nishina effect especially for a hard injection spectrum.

%\end{appendices}

\end{document}